# Fabrication of Soft and Comfortable Pressure-Sensing Shoe Sole for Intuitive Monitoring of Human Quality Gaits

Muhammad Adeel[1], Hasnain Ali[1], Afaque Manzoor Soomro[1,2*], Muhammad Waqas[1]

[1]Department of Electrical Engineering, Sukkur IBA University, Pakistan
[2]Department of Mechanical Engineering, Washington University, USA
*afaquemanzoor@gmail.com

## ABSTRACT

The study discusses the design and fabrication of flexible pressure sensors using Ecoflex/Graphene composites. The fabricated sensor is used for the application of intuitive monitoring of human quality gaits and implementation of the soft and comfortable shoe sole for rehabilitation of the patients with foot disorder is also taken into consideration. The sensor is fabricated using molding and casting technique by sandwiching the thin film Ecoflex/Graphene composites between the copper (Cu) electrodes with the dimension of 15 x 15 mm$^2$ with high sensitivity. There are five pressure sensors integrated in the shoe sole, a sensor at the forefoot, three sensors at the midfoot and one sensor at the lower foot (heel). The behavior of the sensor is negative piezoresistive in which the resistance decreases as the pressure increases. The sensors are embedded in a soft and comfortable shoe sole and then integrated with a laptop or mobile application to monitor and analyze human gait in real-time. Furthermore, a dedicated Graphical User Interface (GUI) is designed to read the data. The pressure sensors are integrated with ESP32 microcontroller which wirelessly transmit data to the GUI and smart phones which could be further used in the intuitive monitoring, rehabilitation of the patients with foot disorder or neuromotor diseases.

**Keywords**: Flexible Pressure Sensor, Human Gait Monitoring, Ecoflex/Graphene Composite, Intuitive Rehabilitation, Real-Time Data Analysis

# INTRODUCTION

Human walking pattern analysis, commonly called gait analysis, offers valuable insights into various aspects of human health and performance. By studying how individuals walk, researchers and clinicians can understand the underlying causes of foot pain, imbalances in gait, and the progression of diseases within the human body. This understanding has applications in clinical diagnosis and research across disciplines such as medicine, biomechanics, physiology, and human performance. Gait analysis involves systematically examining a person's walking pattern, providing critical information for diagnosing conditions, assessing treatment outcomes, and enhancing human performance. One significant tool in gait analysis is the pressure sensor, which plays a pivotal role in understanding the dynamics of human gait and related disorders by embedding them in shoe soles.

The motivation behind this research stems from the imperative need to advance the field of gait analysis and its clinical applications. Foot disorders and neuromotor diseases can significantly impact an individual's quality of life, often leading to pain, discomfort, and impaired mobility. Effective rehabilitation strategies require accurate monitoring of gait patterns to tailor interventions and track progress. Existing gait analysis methods, while informative, may lack real-time insights and comfort for patients. This research aims to bridge this gap by developing flexible pressure sensors using innovative materials, Ecoflex/Graphene composites. These sensors have the potential to revolutionize gait analysis by enabling intuitive monitoring of human gait patterns, leading to improved rehabilitation strategies and personalized treatment plans.

The significance of this study lies in its potential to revolutionize the field of gait analysis and rehabilitation. By designing and fabricating pressure sensors using Ecoflex/Graphene composites and integrating them into a soft and comfortable shoe sole, this research offers an innovative, non-invasive and comfortable solution for monitoring human quality gaits. The user-friendly graphical user interface (GUI) and mobile application provide accessible tools for clinicians, researchers, and patients to monitor gait patterns in real time. Moreover, applying the proposed smart shoe sole for rehabilitation showcases its potential to enhance the quality and effectiveness of treatment for individuals with foot disorders and neuromotor diseases. This study holds promise in improving patient outcomes, advancing research methodologies, and contributing to developing innovative wearable healthcare technologies.

# Human Gaits

Human gait, the intricate coordination of body movements during walking, is a subject of profound scientific interest due to its multifaceted implications for both health assessment and biomechanical research. This phenomenon transcends conventional biomechanics, revealing an intricate interplay of neuromuscular dynamics and motor control that reflect broader health insights [1]. Beyond its locomotive function, gait analysis has evolved into a nuanced diagnostic tool, shedding light on neurological conditions and musculoskeletal anomalies [2].

Wearable sensor systems, notably pressure sensors, have revolutionized the landscape of gait analysis, enabling real-time monitoring of gait patterns [3]. These technologies extend their application beyond mere mobility evaluation to unraveling the underlying physiological and neurological factors governing gait. The integration of these sensors into shoe insoles brings forth an innovative approach, enabling continuous monitoring in diverse environments and contexts [4], [5]. This transition from laboratory-bound assessments to ambulatory monitoring facilitates a holistic understanding of gait dynamics.

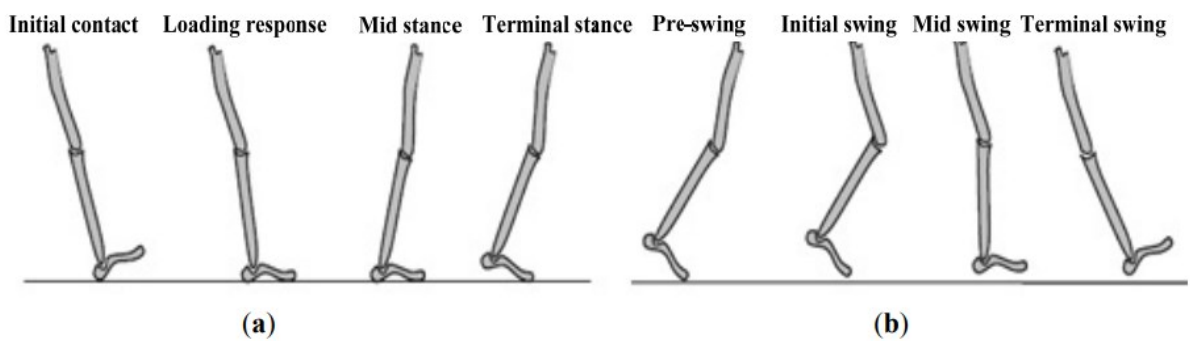

**Fig. 2.1. Gait phases in a normal gait cycle. (a) Gait phases of the stance period; (b) Gait phase of the swing period [6]**

Furthermore, human gait analysis is integral to chronic disease care, offering an avenue for continuous monitoring and proactive intervention. Gait patterns, when assessed comprehensively, unveil subtle deviations that might indicate early stages of diseases such as Parkinson's disease 23[6]. The paradigm shift from episodic assessments to continuous monitoring through wearable sensors positions gait analysis as a potential early warning system for neurological disorders. This aligns with the paradigm of "non-invasive flexible and stretchable wearable sensors with nano-based enhancement for chronic disease care" [6], wherein sensor technologies are harnessed to provide an unobtrusive and real-time health assessment framework.

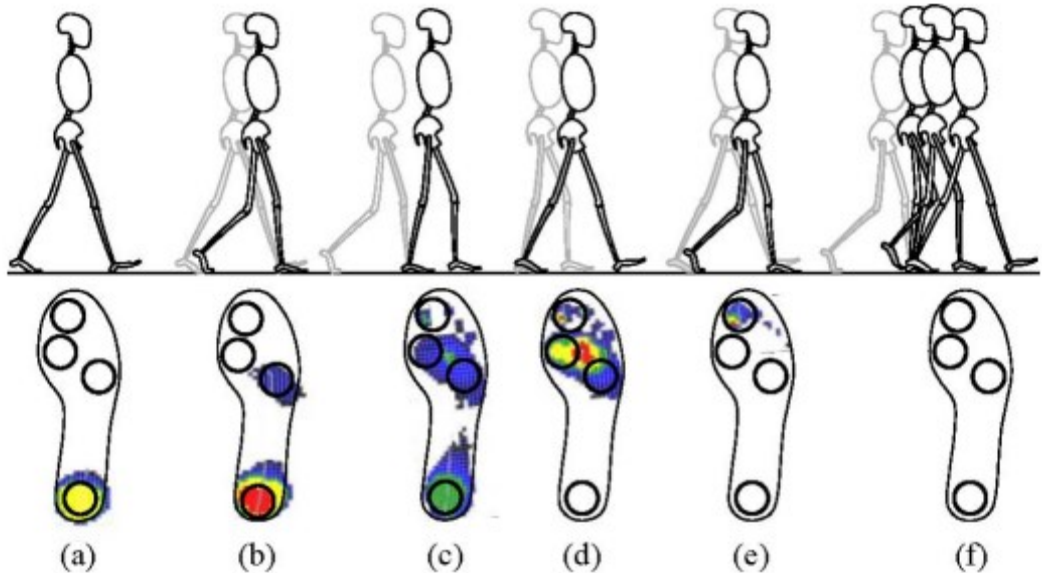

**Fig. 2.2. Fundamental gait phases of a leg (shaded) and their associated GCF patterns in the right foot. (a) Initial contact. (b) Loading response. (c) Mid stance. (d) Terminal stance. (e) Pre-swing. (f) Swing [5], [7], [8]**

In essence, human gait analysis transcends its fundamental role in locomotion, morphing into a dynamic and multifunctional field that leverages advanced sensor technologies to yield insights into neuromuscular dynamics, neurological health, and chronic disease management. The integration of pressure sensors and wearable systems paves the way for personalized and continuous assessment, potentially revolutionizing how we understand and address a myriad of health conditions.

**Neuromotor diseases**

Neuromotor diseases encompass a diverse spectrum of neurological disorders that affect the motor system, leading to disruptions in movement, coordination, and muscle control. Gait analysis emerges as a powerful tool in comprehending the intricacies of these diseases, offering not only diagnostic insights but also opportunities for tailored interventions and rehabilitative strategies [9]. The integration of wearable sensors, particularly pressure sensors, into gait analysis opens avenues for elucidating the manifestations and progression of various neuromotor diseases. These sensors enable continuous monitoring of gait patterns in real-world scenarios, providing a comprehensive picture of movement irregularities that may otherwise go unnoticed during sporadic clinical assessments [2]. This becomes especially significant in conditions like Parkinson's disease, where subtle changes in gait dynamics can be indicative of disease progression [10].

The amalgamation of pressure sensors into wearable insoles offers a non-invasive and unobtrusive means to track gait changes over time [4]. For instance, the work by Kong and Tomizuka underscores the potential of air pressure sensors embedded in shoes to monitor gait patterns and deviations in patients with neuromotor diseases [5]. This approach complements traditional clinical evaluations, providing clinicians with valuable data for informed decision-making and tailored interventions.

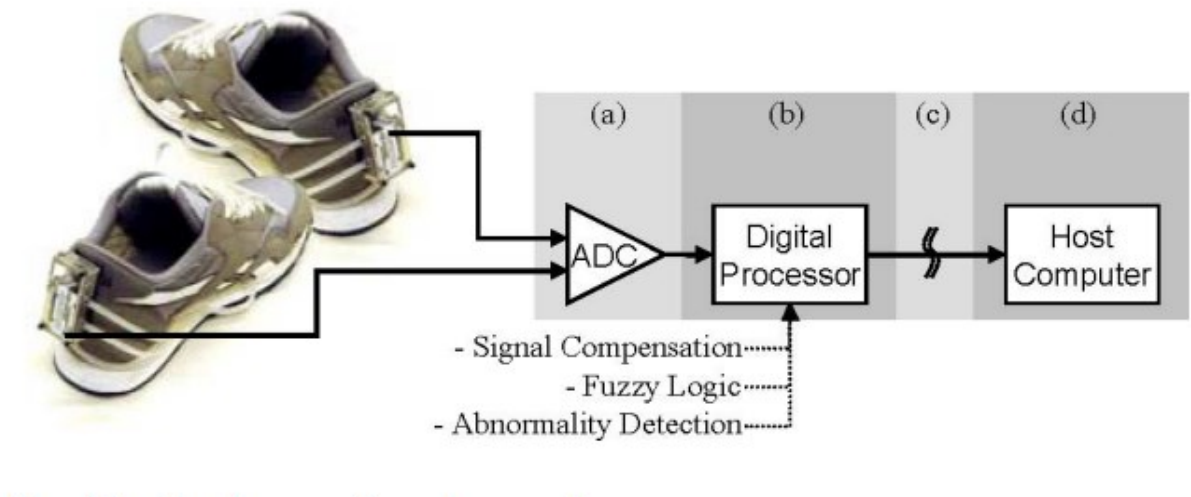

**Fig. 2.3. Implementation of Shoe Soles [5]**

Moreover, advancements in pressure sensor technology facilitate the development of soft prosthetic hands that respond to neural signals, thereby offering improved dexterity and control for individuals with neuromotor impairments [11]. This exemplifies how sensor technologies bridge the gap between biomedical engineering and neurological rehabilitation, forging innovative solutions to enhance the quality of life for those affected by neuromotor diseases.

In conclusion, gait analysis, empowered by wearable pressure sensors, emerges as an indispensable tool for understanding and managing neuromotor diseases. This paradigm shift from episodic assessments to continuous monitoring not only aids in diagnosis but also affords an opportunity for early intervention and personalized rehabilitative strategies. The integration of sensor technologies into wearable systems heralds a new era in neurological care, where data-driven insights and tailored interventions hold the promise of improving the lives of individuals affected by these complex disorders.

## Pressure Sensors

Pressure sensors are pivotal in various fields, ranging from healthcare and biomechanics to industrial applications. These sensors play a critical role in converting mechanical stimuli, such as force or pressure, into measurable electrical signals, enabling the quantification of external pressures or forces acting on a surface [12].

In our research, pressure sensors are instrumental in gait analysis, offering a window into human movement dynamics and the forces exerted during locomotion. These sensors are integrated into soft and comfortable shoe insoles, allowing discreet and continuous monitoring of pressure distribution along the foot while walking [2][5]. The use of pressure sensors in gait analysis facilitates the identification of irregularity in walking patterns, which could indicate various health conditions, including neuromotor diseases [10], [12].

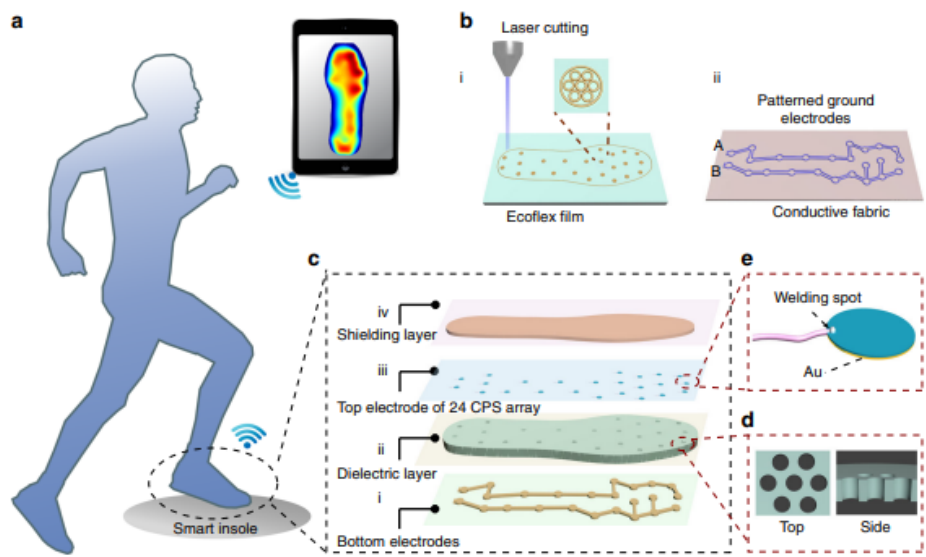

**Fig. 2.4. Schematic illustration of the smart wearable insole system [13]**

Piezoresistive pressure sensors, a subset of pressure sensors, exhibit changes in electrical resistance in response to applied pressure. This property is harnessed to detect and quantify pressure variations [11]. For instance, graphene-based sponge pressure sensors have been developed for rectal model pressure detection [14], [15]. Such sensors offer flexibility, which ensures comfort and adaptability in wearable applications. Additionally, capacitive pressure sensors, another category, rely on changes in capacitance to measure pressure variations [16]. Incorporating these sensors into wearable systems allows for non-invasive and continuous pressure monitoring, aiding in understanding biomechanical processes and health conditions.

Furthermore, advances in nanomaterials like carbon nanotubes (CNTs) and graphene have significantly enhanced the sensitivity and versatility of pressure sensors [17], [18]. Integrating CNTs and graphene into pressure-sensitive composites results in sensors that exhibit high sensitivity, flexibility, and durability, rendering them suitable for wearable applications [19], [20]. This aligns with the trend of developing a "flexible pressure sensor array with multi-channel wireless readout chip" [21] and a "highly compressible integrated supercapacitor-piezo resistance-sensor system for health monitoring" [22], which reflect the ever-evolving landscape of pressure sensor technology.

Pressure sensors are indispensable for gait analysis and biomechanical studies, enabling measuring forces and pressures exerted during human movement. These sensors have witnessed substantial advancements driven by nanomaterial innovations, making them more sensitive, versatile, and adaptable to wearable applications. Integrating pressure sensors into wearable systems offers insights into human motion, health conditions, and biomedical applications, paving the way for enhanced diagnostics, rehabilitation, and personalized healthcare solutions.

**Piezoresistive Pressure Sensor**

Piezoresistive pressure sensors are a class of sensors that utilize changes in electrical resistance in response to applied mechanical pressure. These sensors hold significant relevance in various applications, including biomechanics, healthcare, and industrial sectors, owing to their ability to convert mechanical stimuli into quantifiable electrical signals [23].

Integrating piezoresistive pressure sensors into wearable systems, particularly for gait analysis, presents a novel approach to comprehending human movement dynamics. These sensors are incorporated into shoe insoles, allowing real-time monitoring of pressure distribution during walking [20], [24]. Notably, graphene, a two-dimensional carbon material, has emerged as a pivotal material for enhancing the sensitivity and flexibility of piezoresistive sensors [14], [23]. For instance, Sang et al. developed a graphene-based sponge pressure sensor array for rectal model pressure detection, demonstrating the efficacy of such sensors in capturing subtle pressure variations [14].

The versatility of piezoresistive pressure sensors is further exemplified by their integration with machine learning algorithms, enhancing perception capabilities [11]. This amalgamation of machine learning and sensor technology demonstrates the potential for creating adaptable and intelligent sensor systems for health monitoring and diagnostics [25], [26].

Moreover, piezoresistive pressure sensors find applications beyond gait analysis. They have been utilized for various purposes, such as plantar pressure measurement and gait analysis [3], [20]. Their application extends to monitoring plantar pressure distribution and analyzing biomechanical patterns [27]. This aligns with the broader goal of creating comprehensive capture and analysis systems for skeletal muscle activity during human locomotion [2].

Piezoresistive pressure sensors are instrumental in gait analysis and biomechanical studies, enabling real-time monitoring of pressure distribution and gait dynamics. Incorporating advanced materials like graphene enhances their sensitivity and flexibility, rendering them suitable for wearable applications. With their potential to revolutionize health monitoring, these sensors hold promise for personalized diagnostics, rehabilitation, and chronic disease management, contributing to a data-driven paradigm in healthcare.

## Capacitive pressure sensors

Capacitive pressure sensors constitute a distinct class of sensors that operate on the principle of changes in capacitance resulting from applied mechanical pressure. These sensors hold significant value in various applications, including biomedical and industrial domains, due to their ability to convert mechanical stimuli into measurable electrical signals [16], [28].

The integration of advanced materials, such as polydimethylsiloxane (PDMS) and multi-walled carbon nanotubes (MWCNTs), enhances the sensitivity and performance of capacitive pressure sensors [16], [29]. For instance, a PDMS/MWCNT nanocomposite was utilized for capacitive pressure sensing and electromagnetic interference shielding, demonstrating the versatility of these sensors in multifunctional applications [16]. The conformable nature of PDMS allows for the development of flexible sensor arrays, contributing to the design of more ergonomic wearable systems [30], [31].

Capacitive pressure sensors are crucial in gait analysis and biomechanical studies, providing real-time pressure distribution data during human movement. Integrating advanced materials enhances their sensitivity and flexibility, making them suitable for wearable applications. These sensors contribute to the broader landscape of health monitoring, human-robot interaction, and assistive technologies, opening avenues for personalized diagnostics, rehabilitation, and advancements in human-machine interfaces [32].

## Piezoelectric pressure sensors

Piezoelectric pressure sensors are a specialized category of sensors that leverage the piezoelectric effect to generate an electric charge in response to applied mechanical stress or pressure. This unique property makes them valuable tools in various applications, ranging from biomedical to industrial sectors, where they enable the conversion of mechanical stimuli into measurable electrical signals [33].

The integration of piezoelectric pressure sensors into wearable systems holds immense potential for gait analysis and biomechanical studies. These sensors are adeptly embedded into shoe insoles to capture pressure distribution during human movement, offering insights into gait dynamics [33], [34]. They present a non-invasive means of continuously monitoring pressure changes during locomotion, thereby providing a comprehensive understanding of foot mechanics and biomechanics.

The utilization of piezoelectric materials, such as poly (vinylidene fluoride) (PVDF), enhances the sensitivity and reliability of these sensors [33], [34]. For instance, a piezoelectric pressure sensor array was developed using a PDMS/MWCNT composite, demonstrating its suitability for plantar pressure measurement and electromagnetic interference shielding [16]. Such materials enhance the sensor's ability to detect subtle pressure variations, making them particularly relevant for biomechanical analysis.

## Our Fabricated Pressure Sensor

In gait analysis and biomechanical studies, selecting pressure sensors plays a pivotal role in capturing accurate and reliable pressure distribution data during human movement. Capacitive Pressure Sensors, characterized by their capacity to measure pressure changes through variations in capacitance, have demonstrated their effectiveness in wearable applications, particularly when embedded in shoe insoles. While they offer conformability and real-time monitoring capabilities, their sensitivity is lower than the other types, potentially limiting the detection of subtle pressure changes [35],[36] .

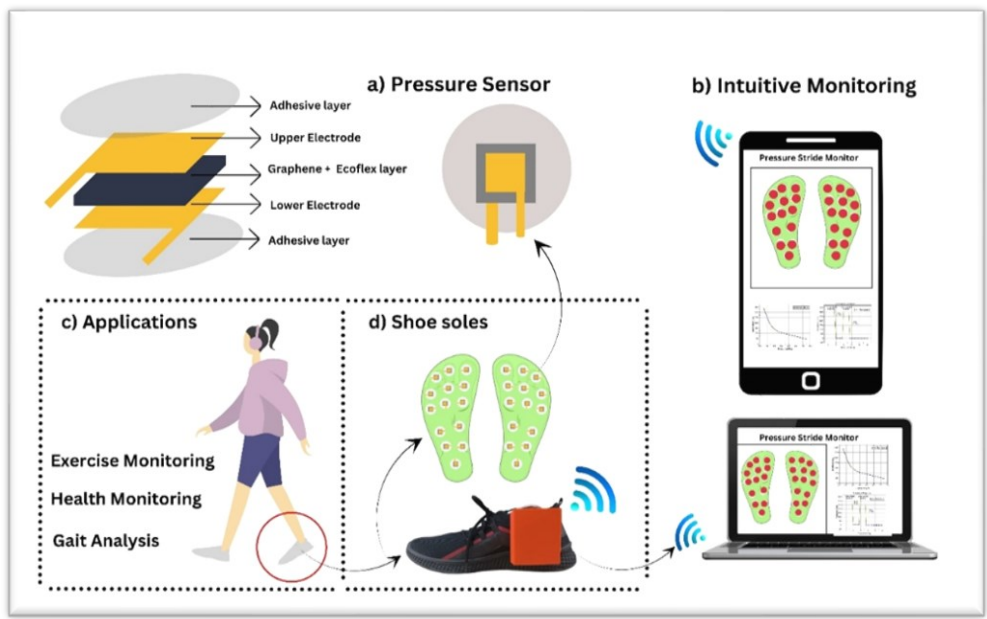

**Fig. 2.5. Flexible Pressure Sensor embedded in shoe soles aims to monitor Human Quality Gaits, which can be used for gait analysis; diagnosis, assessment, or for monitoring the results of treatment; and rehabilitation applications: a) Fabricated pressure sensor. b) Intuitive Monitoring using smart shoe soles. c) applications d) smart shoes with embedded pressure sensors in shoe soles.**

On the other hand, Piezoresistive Pressure Sensors emerge as a strong contender, significantly when fortified with advanced materials like Ecoflex/Graphene composites [20], [33]. These sensors excel in sensitivity due to the piezoresistive effect, which enhances their ability to detect even minute pressure variations during walking [20]. The integration of Ecoflex/Graphene further contributes to flexibility and conformability, aligning well with the ergonomics of wearable systems [20]. Moreover, Piezoresistive sensors enable direct real-time pressure measurements, which is crucial for capturing dynamic gait patterns [23]. The references highlight the successful application of such sensors in various scenarios, including plantar pressure measurement, gait analysis, and even health monitoring for chronic diseases [9], [20].

Piezoelectric Pressure Sensors, renowned for generating electric charges in response to mechanical stress, showcase promising capabilities in capturing pressure distribution during gait [37]. However, while they offer sensitivity, their integration complexity might be comparatively higher, especially for the targeted wearable application [37]. Moreover, the references demonstrate their versatility in applications like tactile sensing in soft prosthetic hands and human-robot interaction [10], [17]. Still, their use in gait analysis could be further explored.

In conclusion, while each type of pressure sensor holds its unique advantages and challenges, the choice of Piezoresistive Pressure Sensors for our project enhanced with Ecoflex/Graphene composites emerges as particularly suitable for intuitive gait monitoring. These sensors exhibit high sensitivity, conformability, flexibility, and real-time monitoring capabilities, as evidenced by successful applications in plantar pressure measurement, gait analysis, and health monitoring [9], [20]. The integration of advanced materials complements the ergonomic requirements of wearables, making them an ideal choice for capturing intricate pressure distribution patterns during human movement.

## Graphene

Graphene, a single layer of carbon atoms arranged in a two-dimensional lattice, stands out as a critical material due to its exceptional electrical, mechanical, and flexible properties [1], [3], [20], [33]. Its high electrical conductivity and sensitivity to mechanical strain make it an excellent candidate for enhancing the performance of pressure sensors [20]. Graphene's integration into pressure sensors, especially with flexible substrates like Ecoflex, contributes to their ability to conform to complex surfaces and capture precise pressure distribution. Its use in fabricating pressure sensors showcases its potential for advancing wearable technologies in healthcare and biomechanics [20], [33], [38], [39].

## Ecoflex

Ecoflex, a flexible and stretchable silicone elastomer, plays a vital role in enhancing the conformability and comfort of pressure sensors. Its mechanical properties allow the sensor to withstand deformation and maintain accurate pressure measurements over time, making it suitable for wearable applications like shoe insoles. The combination of Ecoflex/Graphene results in a composite material that retains sensitivity and flexibility, enabling accurate pressure distribution measurements during gait analysis [20], [33], [40], [41].

### Carbon Nanotubes (CNTs)

CNTs possess excellent electrical and mechanical properties, making them suitable for improving sensor materials' conductivity and mechanical strength [42]. They are often used with other materials like PDMS to create composite materials that can withstand pressure and deformation while maintaining their sensing capabilities [19], [21].

### Polydimethylsiloxane (PDMS)

PDMS is a widely used elastomer known for its flexibility and biocompatibility. It is often used as a substrate or encapsulating material for pressure sensors, providing mechanical support and insulation. When combined with conductive materials like carbon nanotubes, PDMS-based composites can enhance the overall performance and durability of the sensor [19], [21], [43], [44].

### Other Nanocomposites

Various literature mentions the use of other nanocomposites and materials like multi-walled carbon nanotubes (MWCNTs) to improve the conductivity, sensitivity, and mechanical properties of pressure sensors. These materials contribute to the overall effectiveness of the sensor by optimizing its performance in terms of pressure detection and durability [16], [19], [29], [44], [45]

Ecoflex/Graphene composites emerge as superior choices for pressure sensor applications in gait monitoring due to their exceptional combination of properties. Graphene's remarkable electrical conductivity, mechanical strength, and Ecoflex's flexibility and stretchability create a synergistic material blend that accurately captures pressure distribution during human movement. These composites offer the advantage of conforming to complex surfaces, ensuring a comfortable fit for wearable applications such as shoe insoles. Moreover, the integration of Graphene enhances the sensor's sensitivity, allowing them to detect even subtle pressure variations. This unique combination of properties makes Ecoflex/Graphene composites well-suited for the demands of gait analysis, enabling precise, real-time measurements while ensuring user comfort and sensor durability. The references further affirm the success of this composite approach in advancing pressure sensor technology for biomechanical studies and healthcare applications.

### Fabrication Methods

Molding and Casting is a prevalent fabrication technique in the creation of pressure sensors for gait monitoring [8], [20], [22]. This method involves sandwiching thin film Ecoflex/Graphene composites between copper (Cu) electrodes, forming a sensor structure with high sensitivity. The process entails pouring the liquid Ecoflex and Graphene composite into a mold, followed by curing to achieve the desired sensor shape and properties. This method allows for the integration of multiple pressure sensors within the shoe sole, catering to various regions of the foot.

## Electrospinning Machine

Electrospinning is also mentioned in the context of creating nanofibrous scaffolds, which can potentially be utilized in pressure sensors. This technique involves generating nanofibers from a polymer solution using an electric field, resulting in porous structures that can serve as substrates or supporting materials for sensors. While not directly discussed in the context of pressure sensor fabrication in the provided references, electrospinning showcases potential for enhancing sensor performance through novel material designs [46], [47].

## Direct Wire Fabrication

This approach involves precisely depositing material layer by layer to create the desired sensor structure[1], [7] In the context of pressure sensors, it can contribute to creating intricate patterns or arrays of sensing elements, enhancing the sensor's sensitivity and accuracy [1], [24]. While direct-write fabrication is not discussed extensively in the provided references, it remains a relevant method for creating customized and high-resolution pressure sensor arrays.

## Comparison and Conclusion

Molding and casting stand out as the preferred method due to its simplicity, scalability, and ability to create conformable sensors with high sensitivity [21], [23]. It offers the advantage of integrating multiple sensors within a single device, making it well-suited for gait analysis applications. While electrospinning and direct-write fabrication have their merits, they may introduce complexities and challenges, such as equipment requirements and limited conformability [1], [24], [47], [48]. Therefore, considering the balance between ease of fabrication, sensor performance, and practicality, molding and casting using Ecoflex/Graphene composites emerges as the optimal choice for your proposed pressure sensor design in gait monitoring.

# METHODOLOGY

## Fabrication of the Pressure Sensor

Initially, 3D molds of the size 15 x 15 mm$^2$ and the thickness of 1.25mm are designed using SolidWorks 2021 and printed using UltiMaker 3D printer and Cura software. Then the digital scale with division of 0.001g is used to measure Graphene nano-powder having thickness 5-20nm, area size, 10 x 10 um. The measured amount of graphene is 5 mg. The graphene powder is mixed homogenously with Ecoflex-0030, which is commercially provided into two parts, A and B, with the weightage ratio of (1:1) in the 3D printed mold. The homogeneous mixture of the Ecoflex/Graphene is left to cure for 4 hours to prepare a thin film of Ecoflex/Graphene composites. After curing, the thin film of Ecoflex/Graphene composites is sandwiched between two copper (Cu) electrodes, forming a sensor structure with high sensitivity. The sensor is then embedded into the shoes soles and shoe soles are used to make a smart and innovative shoe soles for the intuitive monitoring of human gait analysis. The complete process is illustrated in Fig. 3.1 below.

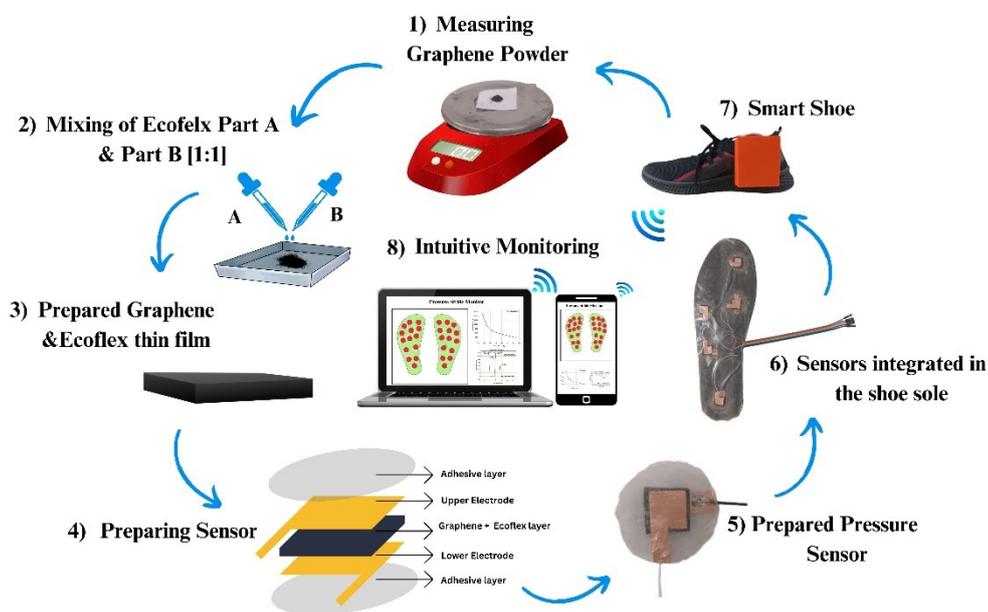

**Fig. 3. 1.** Fabrication and Implementation of Pressure Sensor for monitoring of human quality gaits

## Design of Mold Using SolidWorks

SolidWorks holds significant importance in the Design of molds for pressure sensors, particularly in our context of creating a sensor with specific dimensions and thickness. SolidWorks, a widely-used computer-aided design (CAD) software, offers a range of tools and features that are invaluable for mold design and manufacturing.

In the case of our project, designing a mold for a pressure sensor's thick layer of 15 x 15 mm² and a thickness of 1.25mm using SolidWorks 2021 offers several advantages:

1. **Precision and Accuracy:** SolidWorks enables precise modeling of the mold, ensuring that the dimensions, geometry, and features are accurately represented. This precision is crucial to ensuring the mold corresponds accurately to the intended sensor design.
2. **Visualization:** The software's 3D visualization capabilities allow to visualize the mold in a virtual environment before its physical creation. This helps in identifying potential design flaws, interferences, or other issues that could arise during manufacturing.

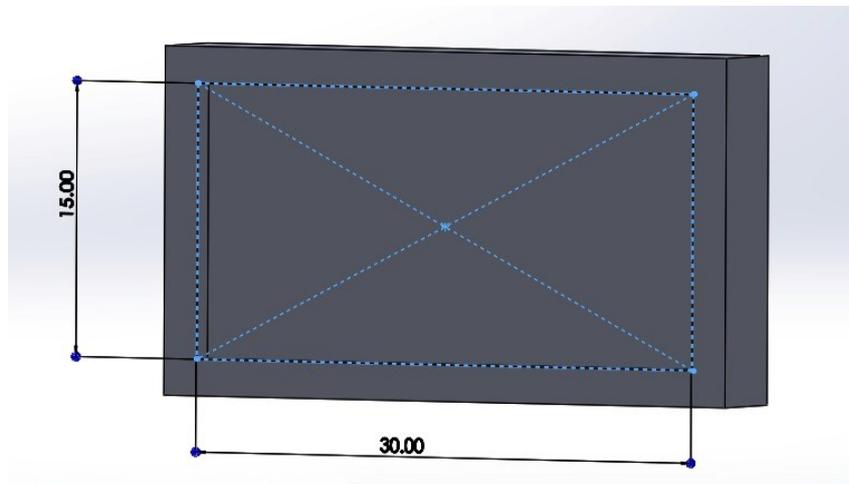

Fig. 3. 2.  Dimensions of the mold

3. **Iterative Design:** SolidWorks facilitates iterative design processes, allowing to make modifications to the mold design as needed quickly. This agility is particularly valuable for refining the mold's geometry to achieve the desired sensor properties.
4. **Interference Detection:** The software's interference detection tools enable the identification of any clashes or overlaps between mold components, ensuring that the mold can be manufactured without any issues.
5. **Simulation:** SolidWorks offers simulation capabilities that allow to assess how the mold will perform under various conditions, helping to predict its behavior during the molding process and optimize its Design accordingly.
6. **Documentation:** SolidWorks aids in the creation of detailed technical drawings and documentation for the mold, which are essential for communicating the Design to manufacturing teams and ensuring accurate fabrication.

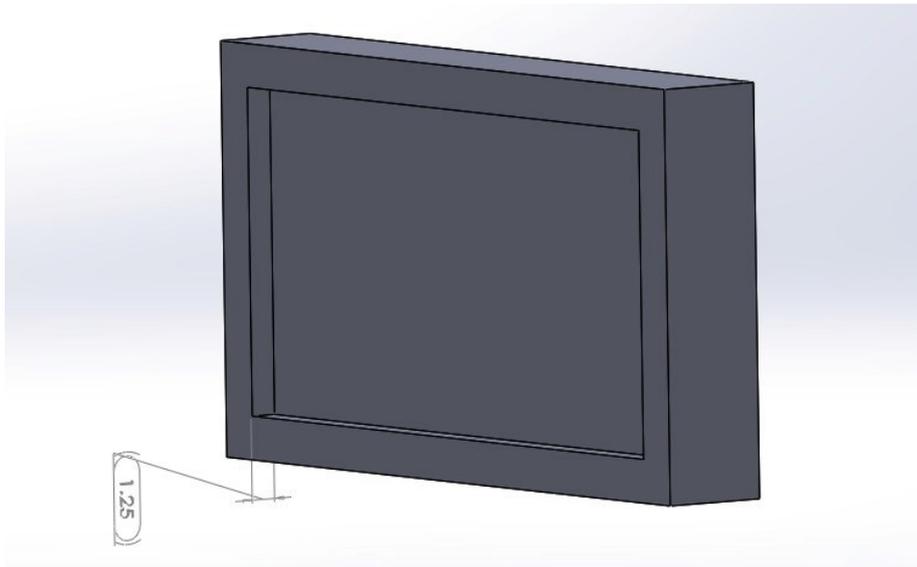

**Fig. 3. 3. Thickness of the mold**

7. **Compatibility:** SolidWorks 2021's compatibility with various manufacturing processes and file formats ensures a seamless transition from the digital mold design to its physical production.

By utilizing SolidWorks to design a mold for our pressure sensor with specific dimensions and thickness, we streamlined the mold fabrication process, minimized potential errors, and ultimately contributed to the successful creation of accurate and functional pressure sensors for our gait monitoring application.

## Homogenization of Ecoflex/Graphene Composites

The process of combining Ecoflex 00-30 and Graphene nano-powder, with a thickness range of 5-20nm and an area size of 10 x 10μm, involves careful and controlled procedures to ensure proper dispersion and incorporation of Graphene into the Ecoflex matrix.

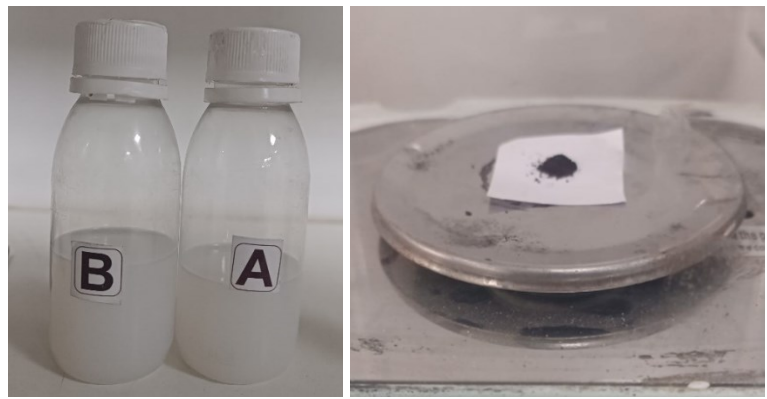

**Fig. 3. 4. Commercial Ecoflex Part A and Part B and Measured Graphene nano-powder**

To achieve this, a predetermined amount of Graphene nano-powder (5 mg) is added to a specified volume of Ecoflex 00-30. The Graphene nano-powder, characterized by its nanoscale thickness and area size, is meticulously weighed to achieve accurate measurements.

The mixture is then subjected to mechanical agitation or other suitable mixing methods. The goal is to ensure uniform dispersion of the Graphene nano-powder within the Ecoflex matrix, allowing for consistent incorporation of Graphene throughout the material.

The resulting composite material, consisting of Ecoflex 00-30 and dispersed Graphene nano-powder, possesses enhanced properties due to the incorporation of Graphene. These properties may include improved mechanical strength, electrical conductivity, and thermal properties, depending on the specific characteristics of the Graphene material used.

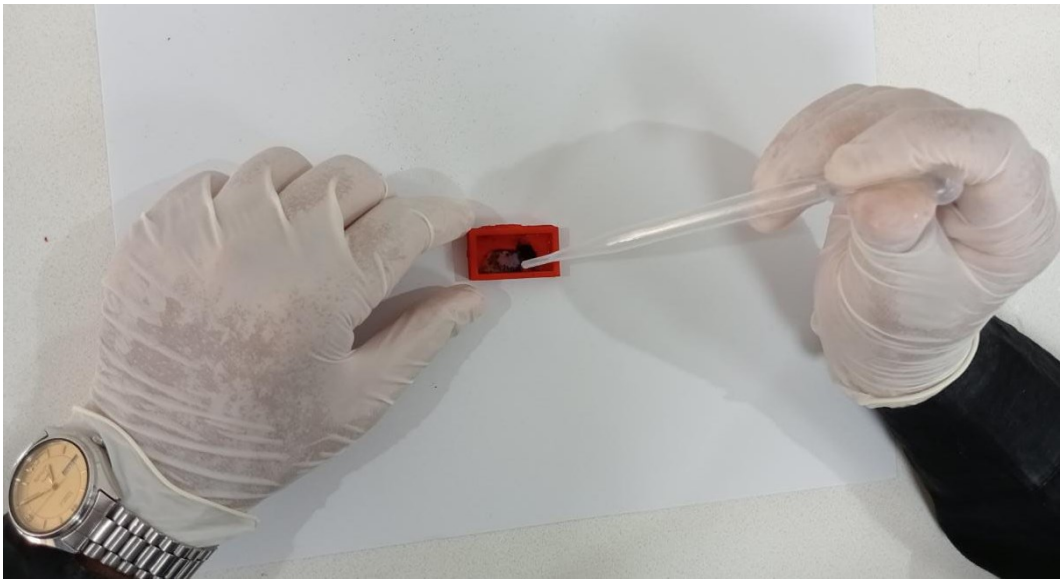

Fig. 3. 5. Homogenization of Ecoflex/Graphene

It's worth noting that the successful homogenization of Ecoflex/Graphene composites requires attention to detail and precision to ensure the desired enhancement of material properties while maintaining the integrity of the final composite structure.

**Fabricated Pressure Sensor**

The fabricated sensor is the outcome of the careful fabrication process involving the integration of Ecoflex/Graphene composites and the subsequent steps taken to create a functional pressure sensor for gait monitoring.

After successfully mixing Ecoflex 00-30 and Graphene nano-powder and attaining a homogenized composite material, the next step involves molding and casting this composite material into a specific sensor design. The designed sensor consists of a 15 x 15 mm² area with a thickness of 1.25 mm. This precise geometry is crucial for achieving consistent sensor performance and accurate pressure measurements.

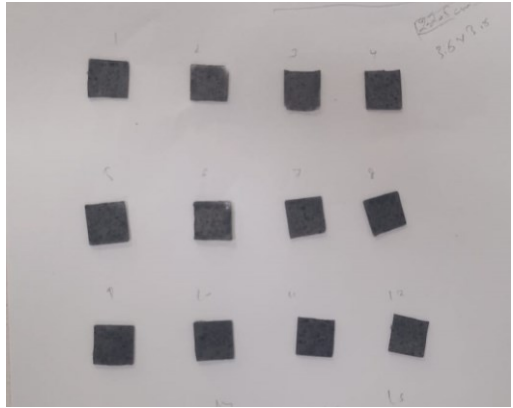

**Fig. 3. 6.  Thin film of Ecoflex/Graphene**

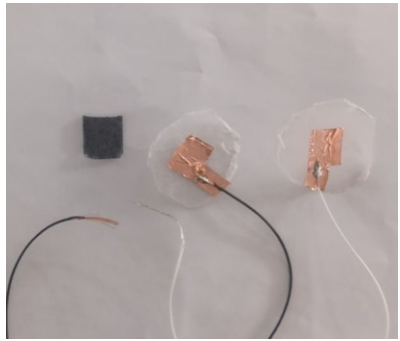

**Fig. 3. 7.  Placing thin film on electrodes of pressure sensor**

During the molding and casting process, the mixed composite material is carefully poured into a mold with the desired sensor shape and dimensions. This mold ensures that the composite material takes on the intended sensor structure while maintaining the uniform distribution of Graphene particles.

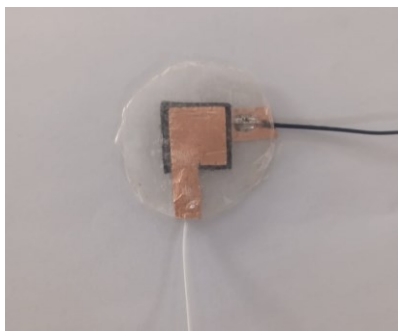

**Fig. 3. 8.  Fabricated Pressure Sensor**

The fabricated pressure sensor exhibits negative piezoresistive behavior, where its electrical resistance changes in response to applied pressure. As pressure is exerted on the sensor, the electrical resistance of the Graphene-Ecoflex composite material undergoes changes, allowing for the detection and quantification of pressure variations.

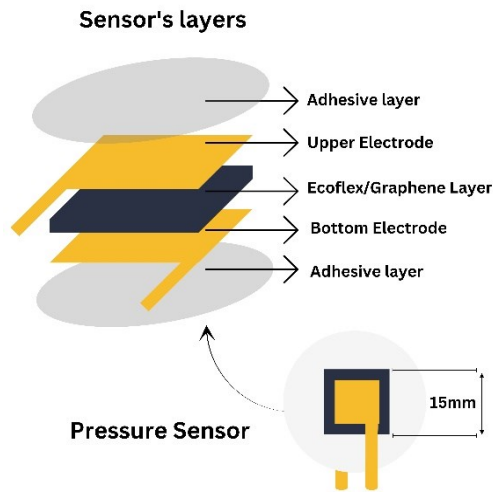

**Fig. 3. 9.  Pressure Sensor's Layers**

Once the material has solidified within the mold, the resulting sensor structure is carefully removed and refined as needed. The sensor's electrical connections, which will facilitate pressure measurements, are established using copper (Cu) electrodes. These electrodes are positioned strategically within the sensor to enable accurate pressure sensing and piezoresistive measurements.

## Integrating Pressure Sensors in Shoe Soles

Integrating pressure sensors into shoe soles significantly advances biomechanical research and clinical applications. In the context of our project, integrating pressure sensors into the shoe soles aims to provide a comprehensive understanding of gait patterns and pressure distribution during walking. This integration involves strategically embedding pressure sensors within the shoe sole structure to capture and analyze the forces exerted by different foot regions during each step [49]. Five pressure sensors have been integrated into the shoe sole design in our project. This configuration enables measuring and analyzing pressure distribution across distinct foot regions. Precisely, one pressure sensor is positioned on the forefoot, allowing for the assessment of pressure dynamics during the initial contact and push-off phase of walking. Additionally, three pressure sensors are strategically located on the midfoot area, providing insights into pressure variations during the midstance and propulsion phases. Lastly, a single pressure sensor is placed on the heel region, enabling the monitoring of pressure changes as the foot undergoes heel strike and early stance.

This sensor arrangement within the shoe sole design is crucial for capturing a comprehensive gait profile.

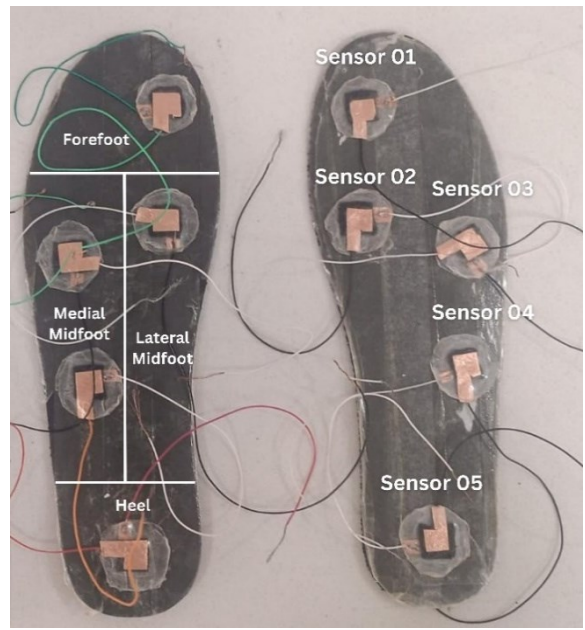

Fig. 3. 10.  Integrating Pressure Sensors in a shoe sole

The data collected from these integrated sensors offer valuable information about foot mechanics, pressure distribution, and potential abnormalities in gait patterns. By analyzing pressure patterns across different foot regions, researchers and clinicians can gain insights into various aspects, such as gait asymmetry, foot posture, and potential foot disorders.

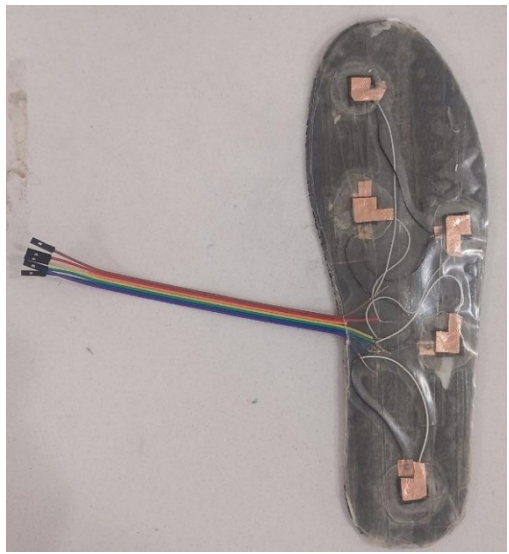

Fig. 3. 11.  Soft and Comfortable Pressure Sensing Shoe Sole

Integrating pressure sensors into shoe soles represents a remarkable intersection of biomechanics, sensor technology, and footwear design. This innovative approach facilitates real-time monitoring of gait dynamics and opens avenues for personalized rehabilitation strategies and early detection of gait-related conditions. Overall, this integration holds immense promise for advancing our understanding of human locomotion and improving patient care in fields related to podiatry, sports science, and clinical rehabilitation.

## Hardware Circuitry

Hardware circuitry plays a pivotal role in enabling the functionality of the pressure sensors and facilitating their integration into the overall monitoring system. We have designed the hardware using following components:

1. 3.7v battery
2. ESP32
3. Resistors
4. Printed Circuit Board (PCB)

In our project, the hardware circuitry is designed to process the changes in electrical resistance exhibited by the piezoresistive pressure sensors when subjected to varying pressures.

The hardware circuitry involves the integration of electronic components to convert the analog changes in resistance into digital signals that can be processed and analyzed. To achieve this, the Voltage Divider Rule (VDR) formula is employed. The VDR formula is used to calculate the output voltage across a voltage divider circuit, which consists of a fixed resistor (R1) and the variable resistance of the piezoresistive sensor (R2).

The VDR formula is given by:

$$V_{out} = V_{in} * \frac{R_2}{R_1 + R_2} \qquad (3.1)$$

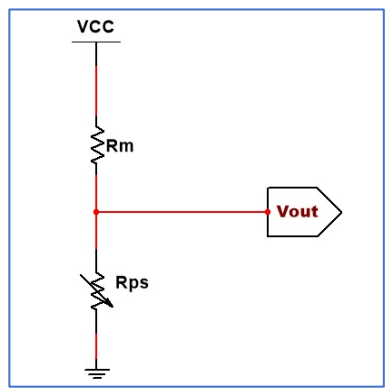

**Fig. 3. 12. Voltage Divider**

In the context of our project, the voltage $V_{in}$ represents the input voltage to the circuit, which is typically supplied by a power source. The variable resistance $R_2$ corresponds to the resistance of the piezoresistive pressure sensor, which changes in response to applied pressure. The fixed resistor $R_1$ serves as a reference element in the voltage divider circuit.

As pressure is exerted on the sensor, its resistance changes, thereby altering the output voltage $V_{out}$ across the circuit. This analog voltage signal is then processed using appropriate signal conditioning techniques, such as amplification and analog-to-digital conversion, to generate a digital signal that can be further analyzed and displayed on a graphical user interface (GUI) or transmitted to a smartphone or computer for real-time monitoring.

By utilizing the VDR formula and the associated voltage divider circuit, the hardware circuitry in our project enables the conversion of mechanical pressure changes into measurable electrical signals. This conversion is a critical step in translating the physical response of the pressure sensors into meaningful data that can be used to monitor and analyze gait dynamics, foot pressure distribution, and related biomechanical parameters.

## Power Supply

The power supply setup is a vital aspect of our sensor system's functionality. In our design, a 3.7V lithium-ion battery has been chosen to power the ESP32 microcontroller. This battery configuration aligns perfectly with the voltage requirements of the ESP32 and facilitates wireless communication between the microcontroller and the Graphical User Interface (GUI) developed using MATLAB. This wireless connection not only allows for data transmission from the pressure sensors embedded in the shoe sole to the GUI but also ensures a user-friendly and unobtrusive experience during gait analysis. The 3.7V battery's compatibility with the ESP32 underscores its role in maintaining consistent and reliable performance throughout the monitoring process.

## Resistors

A resistor is a type of passive electrical component that has two terminals. Its main function is to provide resistance in an electrical circuit, which helps to decrease the current and lower the voltage. Resistors come in a variety of resistance values, from zero ohms all the way up to mega ohms. In our specific circuit, we have chosen to use five resistors with values of 150k ohm connected in series with each sensor.

ESP32 refers to the ESP-WROOM-32, a widely used microcontroller module designed by Espressif Systems. It is known for its integrated Wi-Fi and Bluetooth capabilities, making it suitable for various IoT and wireless communication applications. In our work, the ESP32 microcontroller has been utilized to establish communication between pressure sensors embedded in the shoe soles and the graphical user interface (GUI) we have developed. This microcontroller allows data from the sensors to be collected, processed, and transmitted to the GUI for real-time monitoring and analysis of pressure distribution during gait. [50], [51].

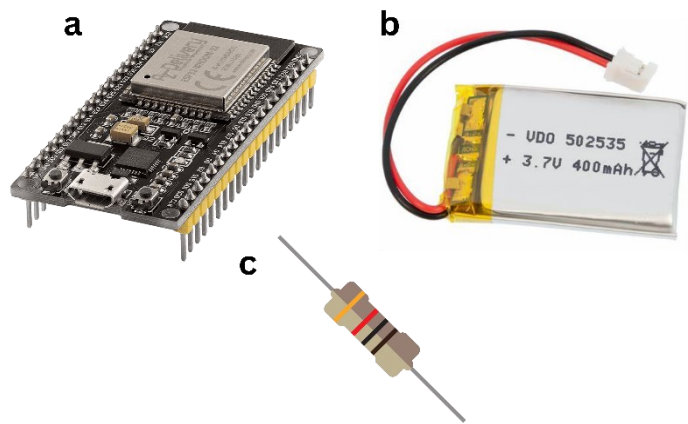

Fig. 3. 13.  a) ESP32 b) 3.7v battery c) a resistor

## Printed Circuit Board

A Printed Circuit Board (PCB) is a mechanical device that connects electrical and electronic components using conductive tracks embedded between layers of non-conductive substrate. Typically, components are soldered onto the PCB, which is used in a wide range of simple electrical and electronic products. PCBs exhibit both electrical and mechanical properties, making them ideal for various applications. The flexibility has been particularly beneficial in medical electronics, where lightweight devices can be manufactured using thin, small-sized flexible and rigid PCBs.

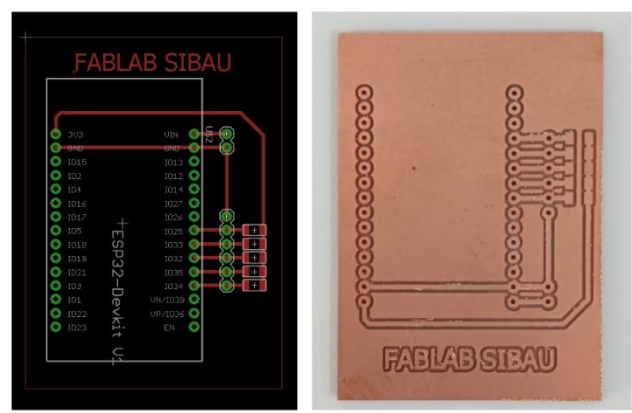

Fig. 3. 14.  Printed Circuit Board (PCB)

The primary materials used in PCBs are copper, solder mask, and nomenclature ink. We utilized Eagle software to design PCB of our circuit, following a three-step process:

i. First, the required circuit is designed in the Eagle software, ensuring proper routing for correct connections.
ii. Then PCB is printed using a 3D printer available in the FabLab of Sukkur IBA University.
iii. Finally, all components are placed and soldered onto their specific locations on the board.

Fig. 3.14 shows the printed circuit board.

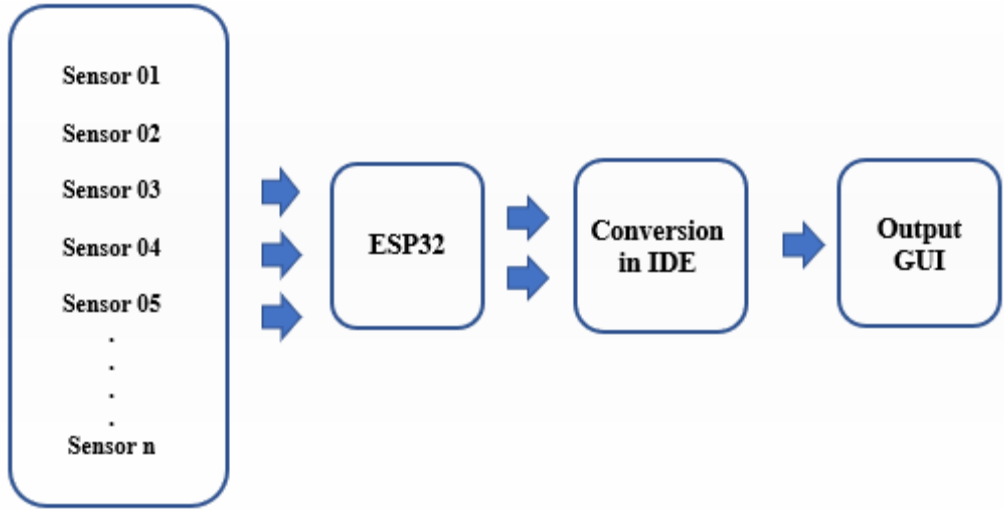

Fig. 3. 15. Data Acquisition of the project

## Arduino

Arduino is an open-source electronics platform that provides a user-friendly environment for creating and prototyping various electronic projects. It includes both hardware and software components, making it accessible for individuals with varying levels of expertise. In our thesis project, Arduino has potentially played a role in interfacing with sensors, collecting data, and possibly assisting in data preprocessing before transmitting it to the GUI. Arduino's versatility and compatibility with various sensors make it a valuable tool in your pressure sensor system.

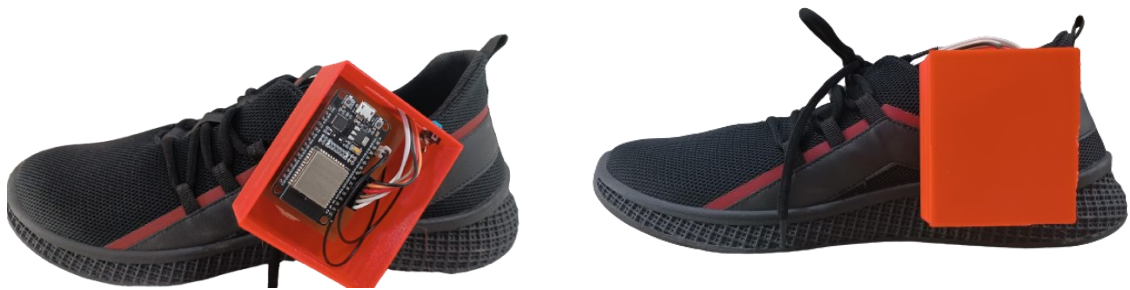

Fig. 3. 16. Smart Shoe

**MATLAB**

MATLAB is a powerful software platform widely used in scientific and engineering fields for data analysis, visualization, and modeling. It offers a range of tools and functions to process and interpret experimental data, making it particularly valuable in our work. You have used MATLAB to receive data transmitted from the ESP32 microcontroller, perform real-time data analysis, visualize pressure distribution patterns, and potentially even apply algorithms to extract meaningful insights from the collected data. Fig. 3.15 shows the data acquisition and integration of the components and the system used for gait analysis. Fig. 3.16 shows the smart shoe having shoe sole embedded in the shoe with the hardware circuitry placed on top of the shoe.

# RESULTS AND DISCUSSION

## Pressure Sensor's Characterizations

Our fabricated pressure sensors are piezoresistive sensors that allow to detect physical pressure, squeezing and weight. The Fig. 4.1 shows the pressure sensor's layers and Fig. 4.2 shows the fabricated pressure sensor.

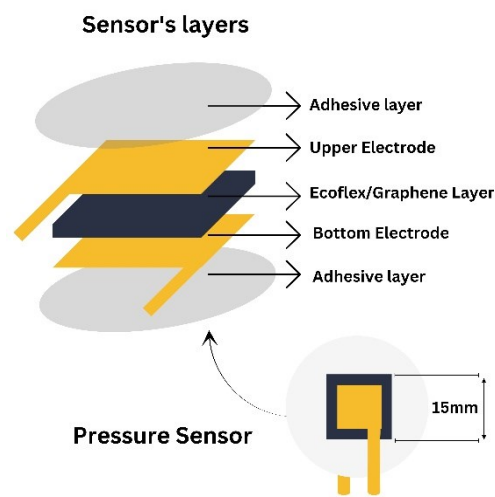

Fig. 4. 1.  Pressure Sensor's Layers

The sensor has active area of 15 x 15 mm$^2$ and thickness of 1.25 mm$^2$. After cyclic period of 50 times, it is deduced that the sensor has infinite resistance when no pressure is applied,150 KΩ after applying light pressure to 200 Ω when maximum pressure is applied.

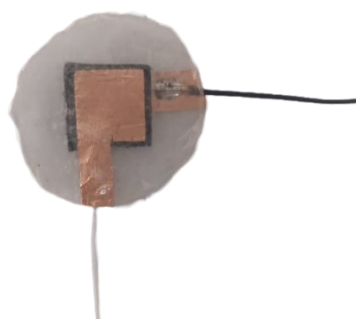

Fig. 4. 2. Fabricated Pressure sensor

There are different ranges of pressures applied on the sensor over the 15 x 15 mm$^2$ surface area where minimum pressure is 0 kPa and maximum applied pressure is 750 kPa. Table 4.1 summarizes pressure sensor's characterizations. Force and pressure conversion is shown in the Table 4.2.

**Table 4.1: Pressure Sensor's Characterizations**

| | |
|---|---|
| Active size area | 15 x 15 mm$^2$ |
| Thickness | 1.25 mm |
| Resistance range | No pressure: Open Circuit |
| | Min. pressure (200kPa): 150KΩ |
| | Max. pressure (750kPa): 200Ω |
| Pressure range | 0 to 750 kPa tested and applied evenly over the 15 x 15 mm$^2$ surface area |
| Power supply | It uses less than 1mA of current (depends on resistor used and supply voltage |
| Sensitivity | 0.02 Pa/ohm |
| Response time | 120 milliseconds (ms) |
| Recovery time | 100 milliseconds (ms) |
| Hysteresis | 6 % |
| Threshold | ± 10 % |
| Cyclic response | 50 |

**Table 4.2: Force to Pressure Conversion**

| Weight (kg) | Force (N) | Pressure (Pa) | Pressure (kPa) |
|---|---|---|---|
| 1 kg | 9.81 N | 43,600 Pa | 43.6 kPa |
| 2 kg | 19.62 N | 87,200 Pa | 87.2 kPa |
| 3 kg | 29.43 N | 130,800 Pa | 130.8 kPa |
| 4 kg | 39.24 N | 174,400 Pa | 174.4 kPa |
| 5 kg | 49.05 N | 218,000 Pa | 218.0 kPa |
| 6 kg | 58.86 N | 261,600 Pa | 261.6 kPa |
| 7 kg | 68.67 N | 305,200 Pa | 305.2 kPa |
| 8 kg | 78.48 N | 348,800 Pa | 348.8 kPa |
| 9 kg | 88.29 N | 392,400 Pa | 392.4 kPa |
| 10 kg | 98.10 N | 436,000 Pa | 436.0kPa |

The force to pressure conversion is done using the following equation:

$$P = \frac{F}{A} \quad (4.1)$$

Where,

F = w * g

g = 9.81 $\frac{m}{s^2}$

The sensitivity value of **0.02 Pa/ohm is** calculated, which highlights that the sensor can detect even subtle variations in pressure, making it suitable for capturing intricate details in human gait patterns. This sensitivity is crucial for accurately monitoring foot movement and providing insights into the quality of gait and potential abnormalities. With a response time of **120 milliseconds**, the pressure sensor demonstrates a quick reaction to pressure alterations. This attribute is vital for real-time monitoring and analysis of gait, enabling the sensor to capture dynamic changes in pressure distribution as a person walks. A recovery time of **100 milliseconds** indicates that the sensor is efficient in returning to baseline conditions. This quality ensures that the sensor is ready for subsequent pressure measurements and minimizes any potential delay in capturing consecutive gait cycles. A **hysteresis of 6%** suggests that the sensor's response remains consistent within this range, which is important for accurate and reliable pressure measurements during gait analysis. The threshold of **± 10%** signifies the allowable range of pressure deviations that the sensor can accurately detect. The sensor has been tested more than 50 times and still maintains its accuracy suggests its robustness and reliability for prolonged use in gait analysis scenarios. All of these specifications collectively underscore the effectiveness of the piezoresistive pressure sensor in our research.

**Testing Pressure Sensor**

The easiest way to determine how pressure sensor works is to connect a multimeter in resistance-measurement mode. Since the resistance changes a lot, an auto-ranging meter works well here.

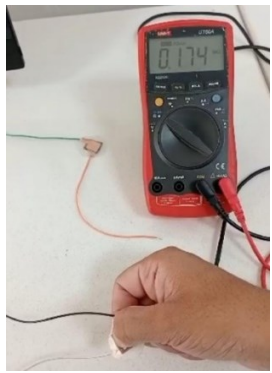

Fig. 4. 3. Measuring the resistance of the pressure sensor

Analog Voltage Reading Method The easiest way to measure a resistive sensor is to connect one end to voltage and the other to a resistor to ground. Then the point between the fixed resistor and the variable pressure sensor is connected to the analog input of a microcontroller such as an Arduino.

**Results**

After careful fabrication of the sensor, different pressures are applied to it to measure Pressure and Resistance values and the behavior of the sensor. The sensor shows piezo-negative behavior, i.e., when pressure is zero, the resistance is observed to be infinite and when a slight change in pressure is observed, the resistance of the sensor decreases.

Table 4.3 presents a time-series record of pressure measurements along with corresponding resistance values. The pressure values remain relatively stable for the initial time points, with consistent resistance values of 3342900 Ohms. However, at the 5th time point, a notable change in pressure is detected, leading to a sudden drop in resistance to 29162.12 Ohms. This drop in resistance suggests that the pressure sensor has responded to a change in external pressure, likely indicating a specific movement or action in the context of human gait.

The subsequent time points continue to show consistent pressure and resistance values, implying a return to the previous pressure condition.

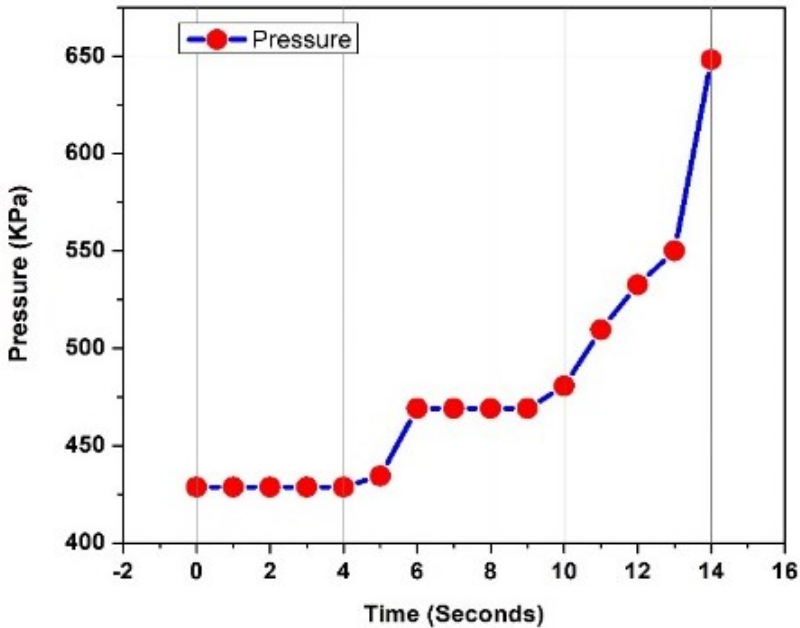

**Fig. 4. 4. Time vs Pressure Graph**

We can observe the behavior using time graphs in Fig. 4.4 and Fig. 4.5.

Table 4.3: Pressure and Resistance Vs Time

| Time (Seconds) | Pressure (Pa) | Resistance (Ohms) |
|---|---|---|
| 0 | 428589.8 | 3342900 |
| 1 | 428589.8 | 3342900 |
| 2 | 428589.8 | 3342900 |
| 3 | 428589.8 | 3342900 |
| 4 | 428589.8 | 3342900 |
| 5 | 434370.1 | 29162.12 |
| 6 | 469052.1 | 29162.12 |
| 7 | 469052.1 | 29162.12 |
| 8 | 469052.1 | 29162.12 |
| 9 | 469052.1 | 29162.12 |
| 10 | 480612.8 | 3342900 |
| 11 | 509514.4 | 3342900 |
| 12 | 532635.8 | 3342900 |
| 13 | 549976.8 | 3342900 |
| 14 | 648242.5 | 8387.898 |

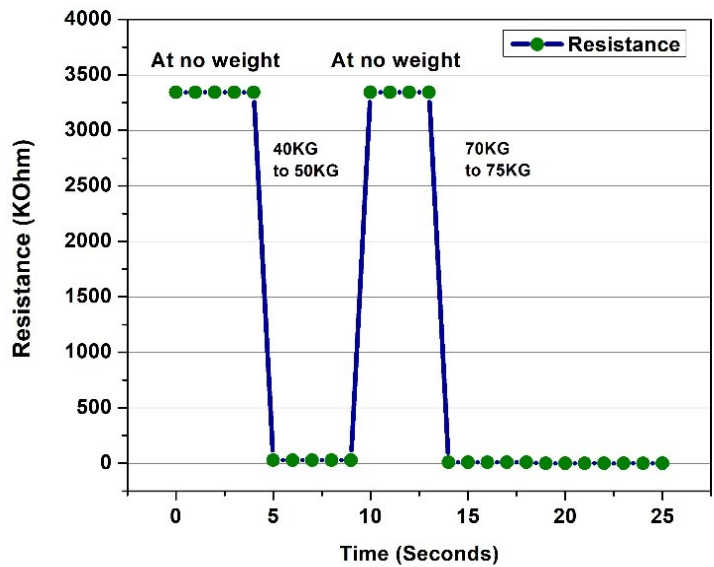

Fig. 4. 5. Time vs Resistance Graph

Table 4.4 provides pressure measurements paired with corresponding resistance values. Similar to the previous table, it highlights the relationship between pressure and resistance. Initially, the pressure remains constant at 428589.8 Pa, resulting in a resistance of 3342900 Ohms. As the pressure increases to 469052.1 Pa, the resistance decreases significantly to 1924700 Ohms. This inverse relationship between pressure and resistance is consistent with the behavior of piezoresistive sensors, where a higher applied pressure leads to a lower resistance due to changes in the material's electrical properties. As pressure further increases or decreases, the resistance values continue to respond accordingly, demonstrating the sensitivity of the sensor to pressure changes.

**Table 4.4: Pressure vs Resistance**

| Pressure (Pa) | Resistance (Ohms) |
|---|---|
| 428589.8 | 3342900 |
| 428589.8 | 3342900 |
| 428589.8 | 3342900 |
| 469052.1 | 1924700 |
| 480612.8 | 1711436.842 |
| 486393.1 | 1620800 |
| 509514.4 | 1333783.333 |
| 532635.8 | 1128771.429 |
| 723386.8 | 463322.9508 |

Similarly, Fig. 4.6 shows the pressure sensor response curve.

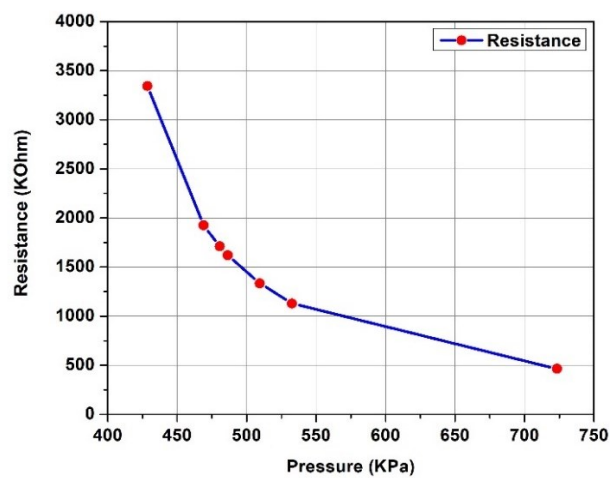

**Fig. 4. 6.  Pressure Response Curve**

Fig. 4.7 shows the comparison of the responsiveness and behavior of our pressure sensor in contrast to a commercial Force Sensing Resistor (FSR) when subjected to varying pressures over time [52]. The fluctuations in resistance reflect the dynamic changes in pressure distribution and the ability of our sensor to accurately capture these changes, which is crucial for gait analysis application. The varying responses and behaviors of the two sensors highlight the unique attributes of our custom sensor compared to the commercial. (FSR).

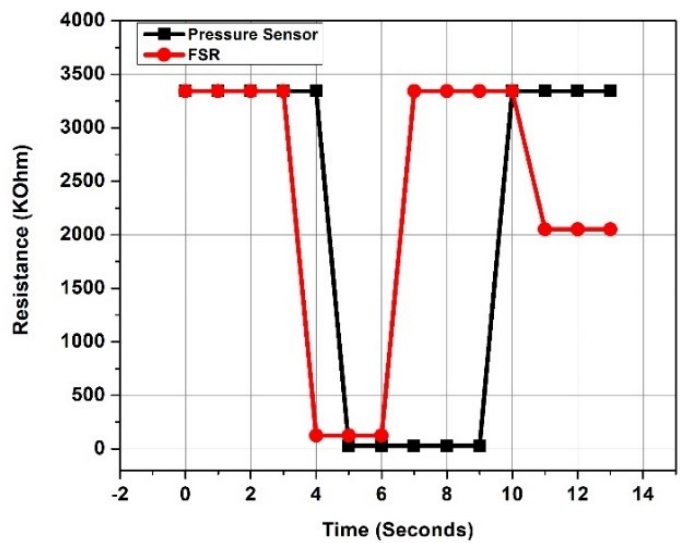

Fig. 4. 7. Comparing Pressure with Commercial FSR

Table 4.5: Comparison of Resistance Responses of Force Sensing Resistor (FSR) and Pressure Sensor over time

| Time (Seconds) | Resistance of Sensor (kohm) | Resistance of FSR (Kohm) |
|---|---|---|
| 0 | 3342.9 | 3342.9 |
| 1 | 3342.9 | 3342.9 |
| 2 | 3342.9 | 3342.9 |
| 3 | 3342.9 | 3342.9 |
| 4 | 3342.9 | 123.81111 |
| 5 | 29.16212 | 123.81111 |
| 6 | 29.16212 | 123.81111 |
| 7 | 29.16212 | 3342.9 |
| 8 | 29.16212 | 3342.9 |
| 9 | 29.16212 | 3342.9 |
| 10 | 3342.9 | 3342.9 |
| 11 | 3342.9 | 2051.325 |
| 12 | 3342.9 | 2051.325 |
| 13 | 3342.9 | 2051.325 |

Overall, our fabricated sensor and the commercial FSR [52] exhibit similar characteristics in terms of size, thickness, pressure range, power supply, sensitivity, response and recovery times, as well as hysteresis and threshold values referring to Table 4.5. These similarities suggest that our sensor holds promise for various applications that demand accurate pressure sensing.

**Graphical User Interface (GUI)**

We've used MATLAB to create a user-friendly Graphical User Interface (GUI) that works seamlessly with the Arduino platform. This GUI is a key part of our sensor system and provides a comprehensive solution for real-time gait analysis.

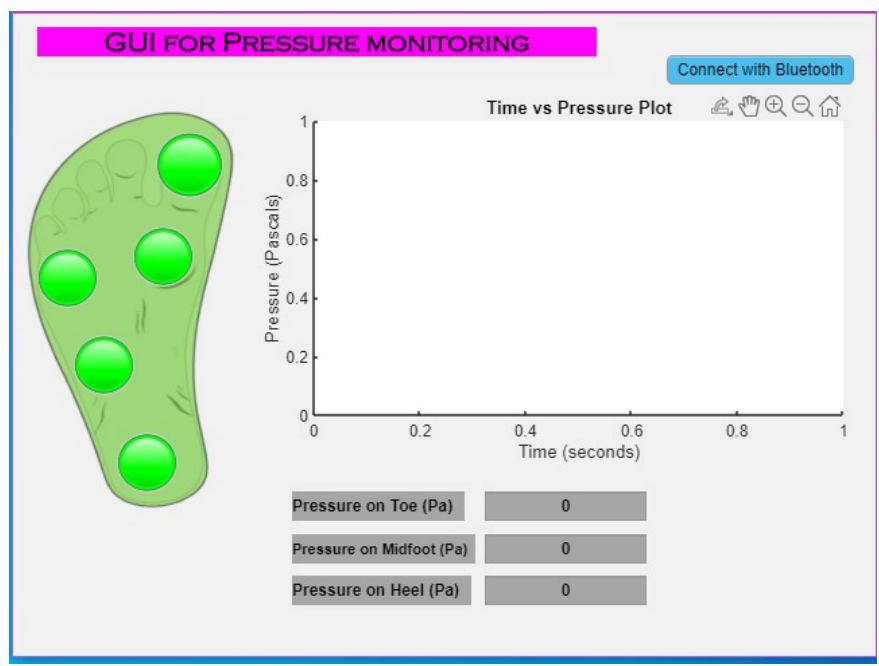

**Fig. 4. 8.  GUI before applying pressure**

The GUI is designed to be easy to use and allows users to visualize and interpret the pressure data collected from the sensors in the shoe soles. It uses the Arduino microcontroller to capture detailed pressure distribution patterns across different areas of the foot - the front, middle, and back.

Users can customize their experience with the GUI, adjusting visualization parameters and focusing on specific areas of interest. The GUI also includes MATLAB's powerful analytical functions, allowing users to dig deeper into the data. They can compare pressure profiles, evaluate statistical parameters related to pressure changes, and monitor gait variations over time.

The GUI communicates in real time with the sensor system, creating a unified ecosystem for comprehensive gait analysis. By combining MATLAB, Arduino, and the GUI, our system provides valuable insights into pressure distribution dynamics during different human gaits. This can help us better understand neurological conditions, track rehabilitation progress, and explore the complex nature of human movement. Figure 4.8 shows the GUI before applying pressure on a shoe sole

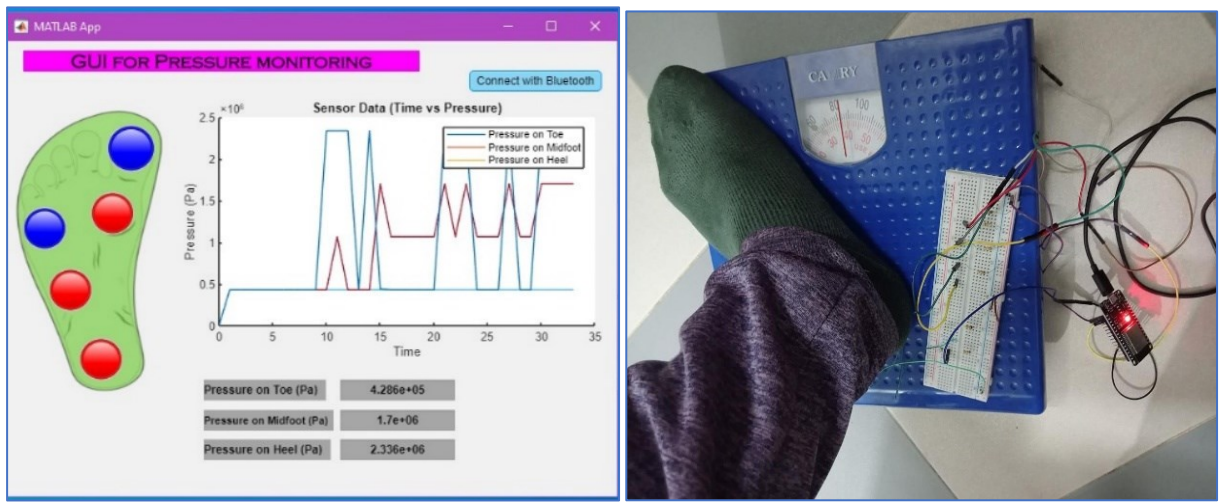

**Fig. 4. 9. GUI after applying pressure**

The GUI reflects the physical layout of the pressure sensors embedded in the shoe sole, showing their positions in real time. When pressure is applied to each sensor, the GUI responds intuitively by changing color intensities - from green for light pressure to red for high pressure, with a progressive change to blue indicating increasing pressure levels. The GUI displays these color-coded changes through a visual representation that clarifies the dynamics of pressure distribution across the shoe sole. Fig. 4.9 Shows the GUI after applying pressure

# FUTURE RECOMMENDATION

As our research project comes to a close, there are numerous opportunities for further study. The rapidly growing field of wearable technology provides an excellent platform for improving the proposed method. The combination of data analytics and machine learning could help us understand complex patterns in the data we've collected, potentially leading to new insights.

In addition, the ongoing advancements in materials science suggest that we could use new sensor materials to improve the sensors' durability, sensitivity, and compatibility with the human body.

It would also be beneficial to integrate the proposed system with comprehensive rehabilitation programs and expand its use to a wider range of neurological disorders. Therefore, the future direction of this research includes improving sensor technology, expanding its applications, and ultimately applying the research findings to benefit patients and the scientific community.

## Conclusion

Our research project has made significant progress in analyzing human walking patterns by incorporating flexible pressure sensors into shoe soles. We evaluated different types of sensors, including piezoresistive, capacitive, and piezoelectric, and found the proposed system highly effective.

We used Ecoflex/Graphene composites to make the sensors, marking a significant change in material science. These materials significantly improved the sensor's sensitivity (0.02 Pa/ohm) and durability (123 milliseconds response time and 100 recovery time).

We strategically placed five sensors in the shoe sole, covering the foot's front, middle, and back. This is a significant step forward in monitoring walking patterns in real-time. It could help us understand standard walking patterns and identify problems early in people with foot or neuromotor disorders.

This project represents a combination of several fields, from material science to sensor engineering, biomechanics, and healthcare. It shows our commitment to increasing knowledge and innovation to improve human health and quality of life. Looking ahead, the proposed system could be helpful in rehabilitation and change how we interact with technology and monitor health. The development of this sensor system and its many applications show the transformative potential of dedicated scientific research in our ever-changing world.

# ACKNOWLEDGEMENT

This undergraduate level final year project was supported by Pakistan Engineering Council and iGNITE.